**Single keyword: XML**

# XML Warehousing and OLAP

Hadj Mahboubi

University of Lyon (ERIC Lyon 2)

5 avenue Pierre Mendès-France, 69676 Bron Cedex, France

Phone: +33 478 773 111 — Fax: +33 478 772 375

hadj.mahboubi@eric.univ-lyon2.fr

Marouane Hachicha

University of Lyon (ERIC Lyon 2)

5 avenue Pierre Mendès-France, 69676 Bron Cedex, France

Phone: +33 478 773 155 — Fax: +33 478 772 375

marouane.hachicha@eric.univ-lyon2.fr

Jérôme Darmont *

University of Lyon (ERIC Lyon 2)

5 avenue Pierre Mendès-France, 69676 Bron Cedex, France

Phone: +33 478 774 403 — Fax: +33 478 772 375

jerome.darmont@eric.univ-lyon2.fr

## INTRODUCTION

With the eXtensible Markup Language (XML) becoming a standard for representing business data (Beyer et al., 2005), a new trend toward XML data warehousing has been emerging for a couple of years, as well as efforts for extending the XQuery language with near On-Line Analytical Processing (OLAP) capabilities (grouping, aggregation, etc.). Though this is not an easy task, these new approaches, techniques and architectures aim at taking specificities of XML into account (e.g., heterogeneous number and order of dimensions or complex measures in facts, ragged dimension hierarchies, etc.) that would be intricate to handle in a relational environment.

The aim of this article is to present an overview of the major XML warehousing approaches from the literature, as well as the existing approaches for performing OLAP analyses over XML data (which is termed XML-OLAP or XOLAP; Wang et al., 2005). We also discuss the issues and future trends in this area and illustrate this topic by presenting the design of a unified, XML data warehouse architecture and a set of XOLAP operators expressed in an XML algebra.

## BACKGROUND

XML warehousing research may be subdivided into three families. The first family focuses on Web data integration for decision-support purposes. However, actual XML warehouse models are not very elaborate. The second family of approaches is explicitly based on classical warehouse logical models (star-like schemas). The third family we identify relates to document warehousing. In addition, recent efforts aim at performing OLAP analyses over XML data.

## XML Web Warehouses

The objective of these approaches is to gather XML Web sources and integrate them into a data warehouse. For instance, Xyleme (2001) is a dynamic warehouse for XML data from the Web that supports query evaluation, change control and data integration. No particular warehouse model is proposed, though.

Golfarelli et al. (2001) propose a semi-automatic approach for building a data mart's conceptual schema from XML sources. The authors show how multidimensional design may be carried out starting directly from XML sources and propose an algorithm for correctly inferring the information needed for data warehousing.

Finally, Vrdoljak et al. (2003) introduce the design of a Web warehouse that originates from XML Schemas describing operational sources. This method consists in preprocessing XML Schemas, in creating and transforming the schema graph, in selecting facts and in creating a logical schema that validates a data warehouse.

## XML Data Warehouses

In his XML-star schema, Pokorný (2002) models a star schema in XML by defining dimension hierarchies as sets of logically connected collections of XML data, and facts as XML data elements.

Hümmer et al. (2003) propose a family of templates enabling the description of a multidimensional structure for integrating several data warehouses into a virtual or federated warehouse. These templates, collectively named XCube, consist of three kinds of XML documents with respect to specific schemas: XCubeSchema stores metadata; XCubeDimension describes dimensions and their hierarchy levels; and XCubeFact stores facts, i.e., measures and the corresponding dimensions.

Rusu et al. (2005) propose a methodology, based on the XQuery technology, for building XML data warehouses, which covers processes such as data cleaning, summarization, intermediating XML documents, updating/linking existing documents and creating fact tables. Facts and dimensions are represented by XML documents built with XQueries.

Park et al. (2005) introduce an XML warehousing framework where every fact and dimension is stored as an XML document. The proposed model features a single repository of XML documents for facts and multiple repositories for dimensions (one per dimension).

Eventually, Boussaïd et al. (2006) propose an XML-based methodology, X-Warehousing, for warehousing complex data (Darmont et al., 2005). They use XML Schema as a modeling language to represent users' analysis needs, which are compared to complex data stored in heterogeneous XML sources. Information needed for building an XML cube is then extracted from these sources.

## XML Document Warehouses

Baril and Bellahsène (2003) envisage XML data warehouses as collections of materialized views represented by XML documents. Views allow filtering and restructuring XML sources, and provide a mediated schema that constitutes a uniform interface for querying the XML data warehouse. Following this approach, the authors have developed the DAWAX system.

Nassis et al. (2005) propose a conceptual approach for designing and building an XML repository, named xFACT. They exploit object-oriented concepts and propose to select dimensions based on user requirements. To enhance the XML data warehouse's expressiveness, these dimensions are represented by XML virtual views. In this approach, the authors assume that all dimensions are part of fact data and that each fact is described in a single XML document.

Rajugan et al. (2005) also propose a view-driven approach for modeling and designing an XML fact repository, named GxFact. GxFact gathers xFACTs (distributed XML warehouses and datamarts) in a global company setting. The authors also provide three design strategies for building and managing GxFact to model further hierarchical dimensions and/or global document warehouses.

Finally, Zhang et al. (2005) propose an approach to materialize XML data warehouses based on frequent query patterns discovered from historical queries. The authors apply a hierarchical clustering technique to merge queries and therefore build the warehouse.

## OLAP Analyses Over XML Data

Chronologically, the first proposals for performing OLAP analyses over XML data mainly rely on the power of relational implementations of OLAP, while more recent research directly relates to XOLAP.

Jensen et al. (2001) propose an integration architecture and a multidimensional UML model for relational and XML data. They also discuss the design of XML databases supporting OLAP analyses. The output of this process is a unified relational representation of data, which are queried with the OQL language. Niemi et al. (2002) follow the same logic, but propose a dedicated language named MDX. In both these approaches, XML data are mapped into a relational database and exploited with relational query languages. Hence, no XML-specific OLAP operator is defined.

Pedersen et al. (2004) also advocate for federating XML data and existing OLAP cubes, but in addition, they propose an algebra composed of three operators. The most fundamental operator in an OLAP-XML federation, decoration, attaches a new dimension to a cube with respect to linked XML elements. Selection and generalized projection help filter and aggregate fact measures, respectively. They more or less correspond to the classical OLAP slice and dice operators. These three operators are implemented with an extension of the $SQL_M$ language, $SQL_{XM}$, which helps associate XPath queries to $SQL_M$ queries. $SQL_M$ is itself an extension of SQL for processing multidimensional data.

Eventually, Park et al. (2005) propose an OLAP framework for XML documents called XML-OLAP and introduce the notion of XML cube (XQ-Cube). A specific multidimensional language (XML-MDX) is applied on XQ-Cubes. Wang et al. (2005) also propose a general aggregation operator for XML, GXaggregation, which forms the base of XCube, an extension of the traditional cube operator for XML data. These two operators have been implemented as an extension of XQuery. In opposition, Wiwatwattana et al. (2007) argue that such an extension from the relational model cannot address the specific issues posed by the flexibility of XML. Hence, they propose an XML warehouse-specific cube lattice definition, a cube operator named X^3, and a generalized specification mechanism. They also discuss the issue of cube computation and compare several alternative algorithms. X^3 has been implemented in C++ within the TIMBER XML-native Database Management System (DBMS).

## UNIFIED XML WAREHOUSING AND XOLAP MODELS

### XML Warehouse Architecture

Previous XML warehousing approaches assume that the warehouse is composed of XML documents that represent both facts and dimensions. All these studies more or less converge toward a unified XML warehouse model. They mostly differ in the way dimensions are handled and the number of XML documents that are used to store facts and dimensions.

A performance evaluation study of these different representations has been performed by Boukraa et al. (2006). It showed that representing facts in one single XML document and each dimension in one XML document allowed the best performance.

Moreover, this representation also allows to model constellation schemas without duplicating dimension information. Several fact documents can indeed share the same dimensions. Also, since each dimension and its hierarchy levels are stored in one XML document, dimension updates are more easily and efficiently performed than if dimensions were either embedded with the facts or all stored in one single document.

Hence, we propose to adopt this architecture model to represent XML data warehouses. It is actually the translation of a classical snowflake schema into XML. More precisely, our reference data warehouse is composed of the following XML documents:

1. *dw-model.xml* represents the warehouse metadata (basically the warehouse schema);
2. *facts.xml* helps store the facts, i.e., dimension identifiers and measure values;
3. *dimension$_d$.xml* helps store a given dimension *d*'s attribute values.

Figure 1 represents the *dw-model.xml* document's graph structure. *dw-model.xml* defines the multidimensional structure of the warehouse. Its root node, *DW-model*, is composed of two types of nodes: *dimension* and *FactDoc*. A *dimension* node defines one dimension, its possible hierarchical levels (*Level* elements) and attributes (including their types), as well as the path to the corresponding *dimension$_d$.xml* document. A *FactDoc* element defines a fact, i.e., its measures, internal references to the corresponding dimensions, and the path to the *facts.xml* document. Note that several *FactDoc* elements are possible, thus enabling constellation schemas.

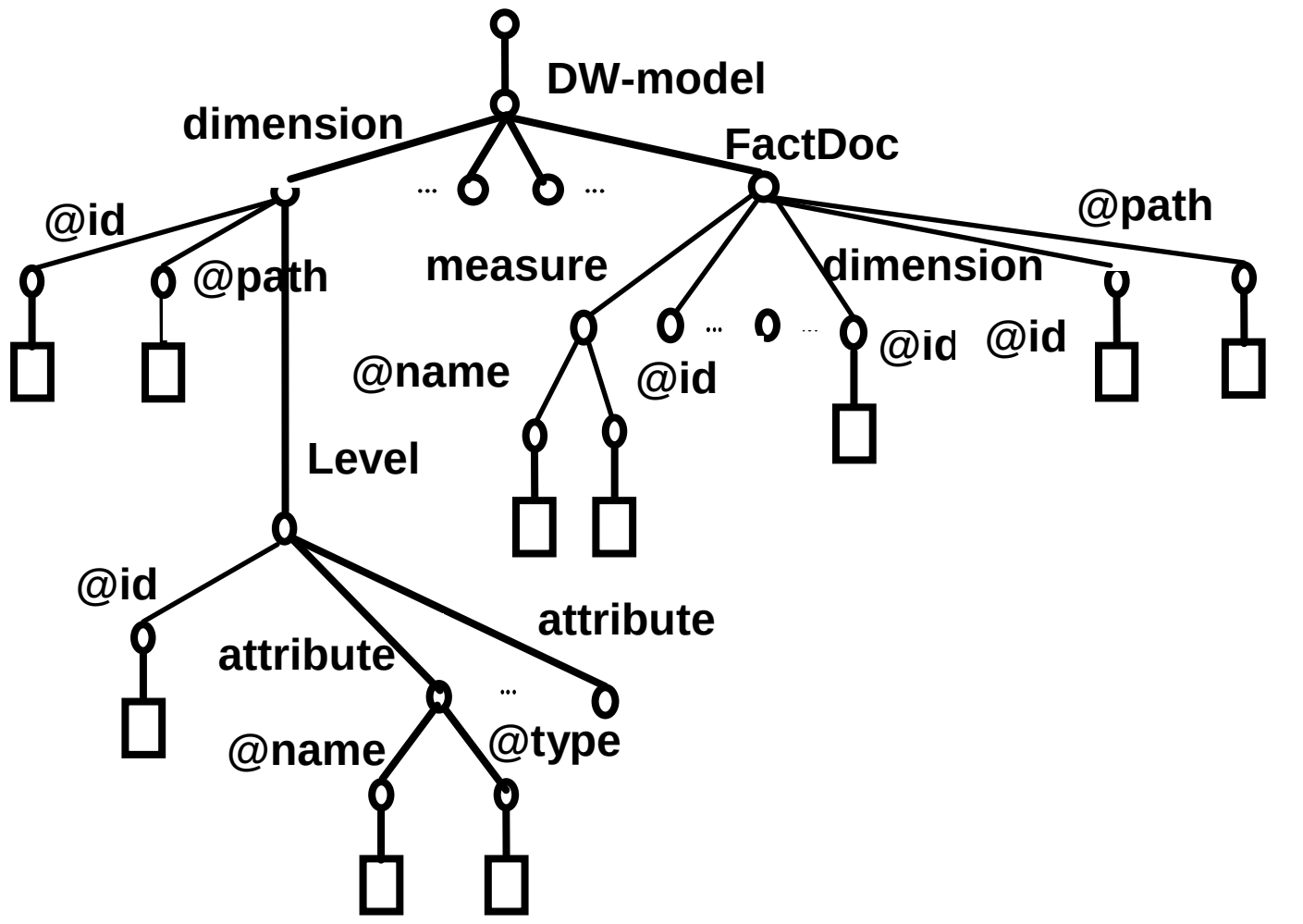

Figure 1: *dw-model.xml* graph structure

Figure 2 represents the *facts.xml* document's graph structure. *facts.xml* stores the facts and is composed of *fact* nodes defining measures and dimension references. The document root node, *FactDoc*, is composed of *fact* sub-elements, each of whose instantiates a fact, i.e., measure values and dimension references. These identifier-based references support the fact-to-dimension relationship.

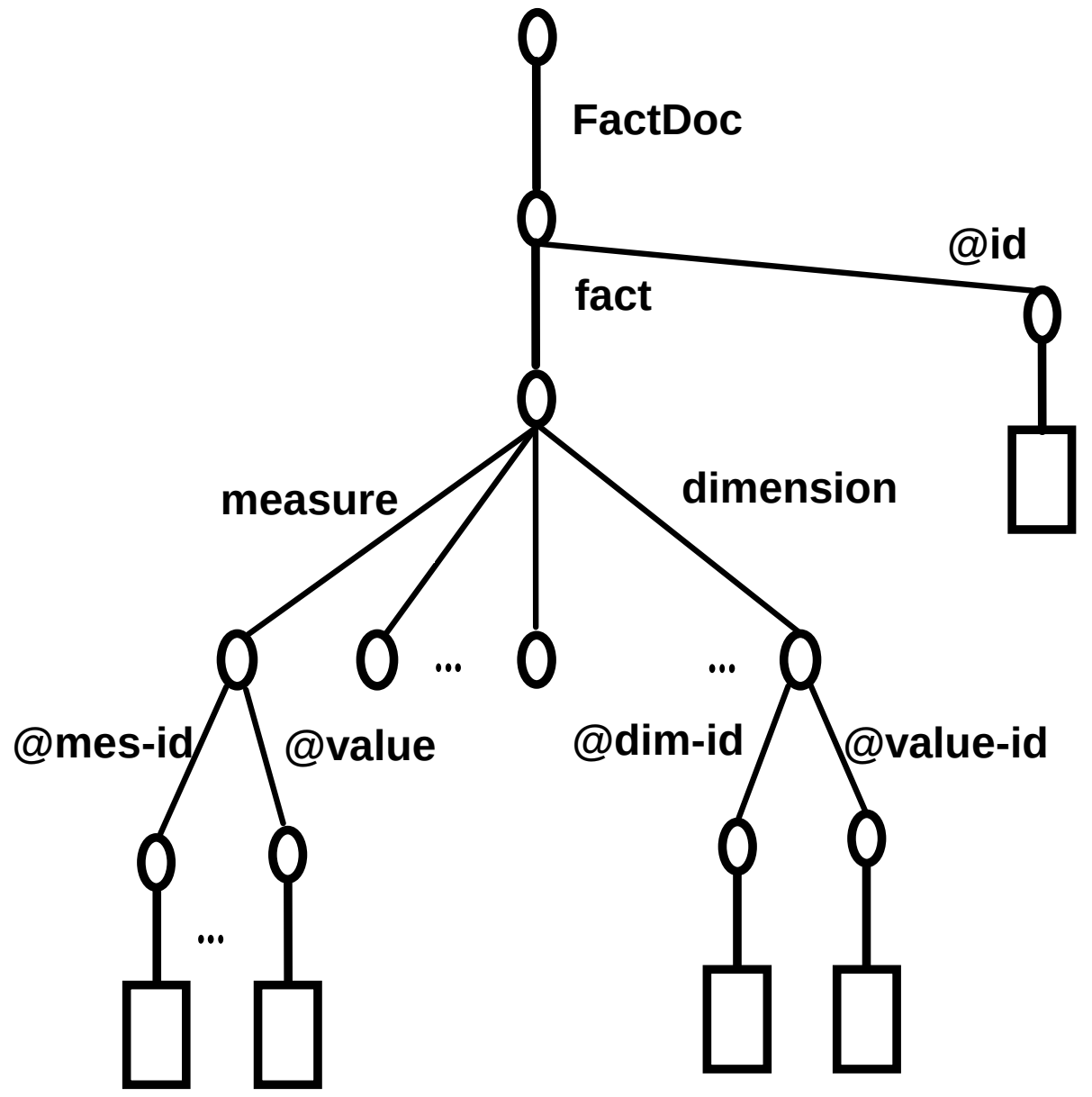

Figure 2: *facts.xml* graph structure

Finally, Figure 3 represents the *dimension$_d$.xml* document's graph structure. *dimension$_d$.xml* helps instantiate dimension *d*, including any hierarchy level. Its root node, *dimension*, is composed of *Level* nodes. Each one defines a hierarchy level composed of *instance* nodes that each defines the level's attribute values. In addition, an *instance* element contains *Roll-up* and *Drill-down* attributes that define the hierarchical relationship within dimension *d*.

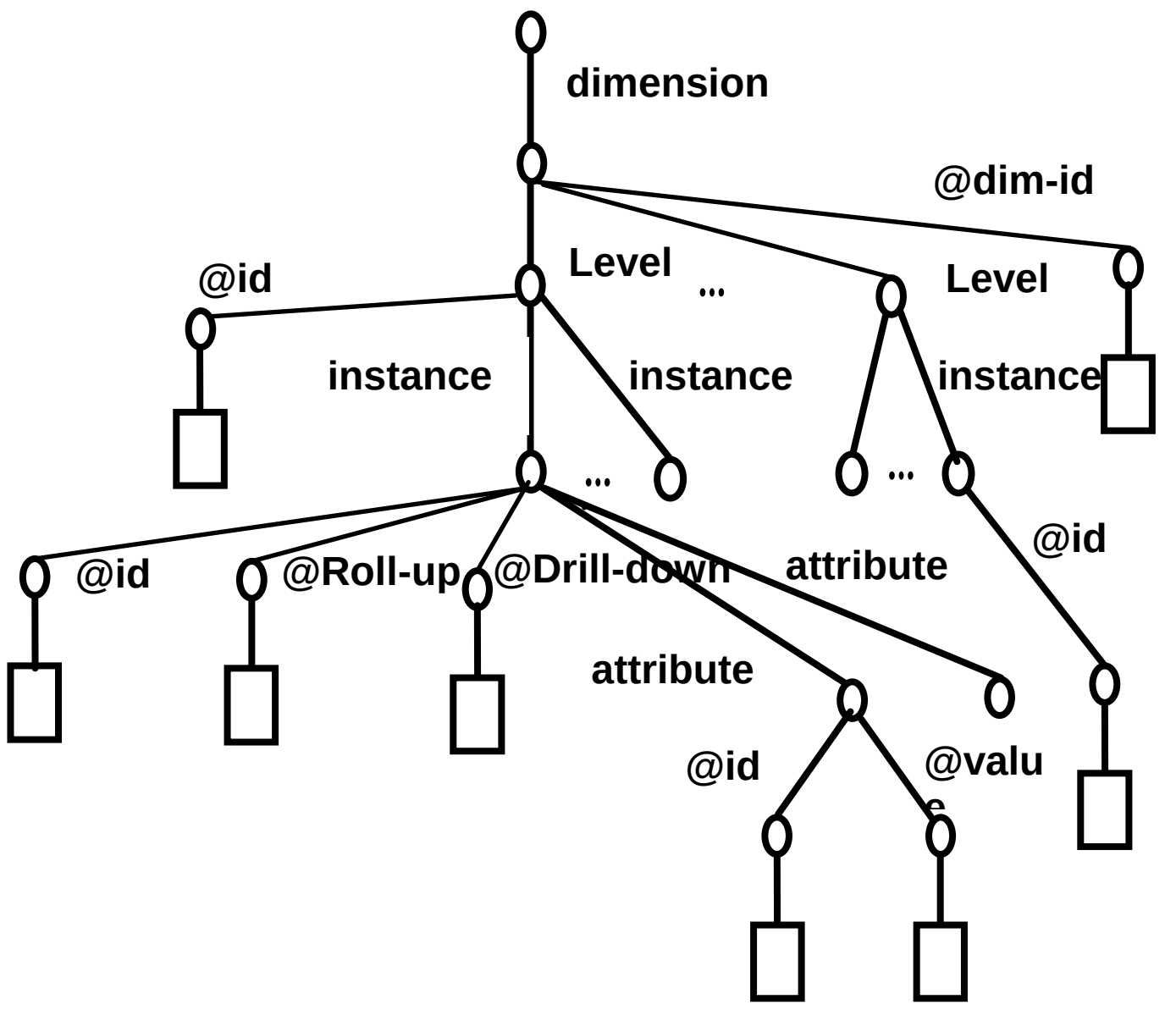

Figure 3: *dimension$_d$.xml* graph structure

Algebraic expression of XOLAP operators

These last decade's efforts for formalizing OLAP algebras (Agrawal et al., 1997; Gyssens and Lakshmanan, 1997; Thomas and Datta, 2001; Ravat et al., 2006) have helped design a formal framework and well-identified operators. Existing OLAP operators, previously defined in either a relational or multidimensional context, must now be adapted to the data model of XML documents (namely trees or, more generally, graphs) and enriched with XML-specific operators.

Existing approaches that aim at XOLAP do not fully satisfy these objectives. Some favor the translation of XML data cubes in relational, and query them with extensions of the SQL language. Others tend toward multidimensional solutions that exploit XML query languages such as XQuery or XML-MDX. However, in terms of algebra, these works only propose a fairly limited number of operators.

As Wiwatwattana et al. (2007), we aim at a native-XML solution that exploits XQuery. As a first step toward an XOLAP platform, we initiated a previously inexistent formal framework in the XML context by demonstrating how the TAX Tree Algebra for XML (Jagadish et al., 2001) could support OLAP operators. Among the many XML algebras from the literature, we selected TAX for its richness. TAX indeed includes, under its logical and physical forms, more than twenty operators, which allows us many combinations for expressing XOLAP operators. Furthermore, TAX's expressivity is widely acknowledged, since this algebra can be expressed with most XML query languages, and especially XQuery, which we particularly target because of its standard status. Finally, TAX and its derived algebra TLC (Paparizos et al., 2004) provide a query optimization framework that we can exploit in the future, since performance is a major concern when designing decision-support applications that are integrally based on XML and XQuery.

We expressed in TAX the main usual OLAP operators: cube, rotate, switch, roll-up, drill-down, slice, dice, pull and push. By doing so, we significantly expanded the number of available XOLAP operators, since up to now, related papers only proposed at most three operators each (always including the cube operator). We have also implemented these XOLAP operators into a software prototype that helps generate the corresponding XQuery code. This querying interface is currently coupled to TIMBER XML-native DBMS, but it is actually independent and could operate onto any other DBMS supporting XQuery.

## FUTURE TRENDS

Historically, XML data warehousing and OLAP approaches very often adapted existing, efficient relational solutions to the XML context. This was a sensible first step, but new approaches must now (and have definitely started to) take the specificities of XML data into account. More specifically, XML warehousing approaches now have to handle heterogeneous facts that may be described each by different dimension sets and/or hierarchy levels. Moreover, because of the variety of data sources, some fact and dimension data may also be missing. Finally, and especially when dealing with complex data such as multimedia data, measures may be non-numerical. For instance, if the observed fact is the health status of a patient, a radiography could be a measure.

Regarding XOLAP, let us take on an algorithmic metaphor. Most authors, and especially Wiwatwattana et al. (2007) and their excellent X^3 operator, have adopted a depth-first approach by fully developing one operator only, which is arguably of little use alone. On the other hand, we have adopted a breadth-first approach by proposing a wider range of operators that, however, only apply onto quite “regular” XML data. Both approaches should aim at completion, in breadth and depth, respectively, to achieve a full XOLAP environment; and are in our opinion complementary. In the next step of our work, we will indeed take inspiration from X^3’s principle to enhance the rest of our XOLAP operators and truly make them XML-specific. For instance, we are actually currently working on performing roll-up and drill-down operations onto the ragged hierarchies defined by Beyer et al. (2005).

## CONCLUSION

More than a mere fashion related to XML's popularity, XML data warehousing and OLAP come from a true need. The complexity of data warehousing and OLAP technologies indeed makes them unattractive to many potential users and there is a growing need in the industry for simple, user-friendly Web-based interfaces, which is acknowledged by decision support system vendors (Lawton, 2006). Hence, though obvious and important difficulties do exist, including the maturity of XML-native DBMSs, especially regarding performance, we think that XML data warehousing and OLAP are promising approaches, and the best able to handle the numerous, so-called complex data (Darmont et al., 2005) that flourish on the Web, in a decision-support context.

## REFERENCES

Agrawal, R., Gupta, A., & Sarawagi, S. (1997). Modeling Multidimensional Databases. In *13th International Conference on Data Engineering (ICDE 97), Birmingham, UK* (pp. 232-243). Los Alamitos: IEEE Computer Society.

Baril, X., & Bellahsène, Z. (2003). Designing and Managing an XML Warehouse. In Chaudhri, A. B., Rashid, A., & Zicari, R. (Eds.), *XML Data Management: Native XML and XML-enabled Database Systems* (pp. 455-473). Boston: Addison Wesley.

Beyer, K. S., Chamberlin, D. D., Colby, L. S., Özcan, F., Pirahesh, H., & Xu, Y. (2005). Extending XQuery for Analytics. In *2005 ACM SIGMOD International Conference on Management of Data (SIGMOD 05), Baltimore, USA* (pp. 503-514). New York: ACM Press.

Boukraa, D., Ben Messaoud, R., & Boussaïd, O. (2006). Proposition d'un Modèle physique pour les entrepôts XML. In *Atelier Systèmes Décisionnels (ASD 06), 9th Maghrebian Conference on Information Technologies (MCSEAI 06), Agadir, Morocco.* Agadir: MIPS-Maroc.

Boussaïd, O., Ben Messaoud, R., Choquet, R., & Anthoard, S., (2006). X-Warehousing: An XML-Based Approach for Warehousing Complex Data. In *10th East-European Conference on Advances in Databases and Information Systems (ADBIS 06), Thessaloniki, Greece* (pp. 39-54); *Lecture Notes in Computer Science,* 4152. Berlin: Springer.

Cheng, K., Kambayashi, Y., Lee, S. T., & Mohania, M. K. (2000). Functions of a Web Warehouse. In *Kyoto International Conference on Digital Libraries 2000* (pp. 372-379). Kyoto: Kyoto University.

Darmont, J., Boussaïd, O., Ralaivao, J. C., & Aouiche, K. (2005). An Architecture Framework for Complex Data Warehouses. In *7th International Conference on Enterprise Information Systems (ICEIS 05), Miami, USA* (pp. 370-373). Setúbal: INSTICC.

Golfarelli, M., Rizzi, S., & Vrdoljak, B. (2001). Data Warehouse Design from XML Sources. In *4th International Workshop on Data Warehousing and OLAP (DOLAP 01), Atlanta, USA* (pp. 40-47). New York: ACM Press.

Gyssens, M., & Lakshmanan, L. V. S. (1997). A Foundation for Multi-dimensional Databases. In *23rd International Conference on Very Large Data Bases (VLDB 97), Athens, Greece* (pp. 106-115). San Francisco: Morgan Kaufmann.

Hümmer, W., Bauer, A., & Harde, G. (2003). XCube: XML for data warehouses. In *6th International Workshop on Data Warehousing and OLAP (DOLAP 03), New Orleans, USA* (pp. 33-40). New York: ACM Press.

Jagadish, H. N., Lakshmanan, L. S. V., Srivastava, D., & Thompson, K. (2001). TAX: A Tree Algebra for XML. In *8th International Workshop on Database Programming Languages (DBPL 01), Frascati, Italy* (pp. 149-164). *Lecture Notes in Computer Science,* 2397. Berlin: Springer.

Jensen, M. R., Moller, T. H., & Pedersen, T. B. (2001). Specifying OLAP cubes on XML data. *Journal of Intelligent Information Systems,* 17(2-3), 255-280.

Lawton, G. (2006). Making Business Intelligence More Useful. *Computer,* 39(9), 14-16.

Niemi, T., Niinimäki, M., Nummenmaa, J., & Thanisch, P. (2002). Constructing an OLAP cube from distributed XML data. In *5th International Workshop on Data Warehousing and OLAP (DOLAP 02), McLean, USA* (pp. 22-27). New York: ACM Press.

Nassis, V., Rajugan, R., Dillon, T. S., & Rahayu, J. W. (2005). Conceptual and Systematic Design Approach for XML Document Warehouses. *International Journal of Data Warehousing & Mining,* 1(3), 63-86.

Pedersen, D., Pedersen, J., & Pedersen, T. B. (2004). Integrating XML Data in the TARGIT OLAP System. In *20th International Conference on Data Engineering (ICDE 04), Boston, USA* (pp. 778-781). Los Alamitos: IEEE Computer Society.

Paparizos, S., Wu, Y., Lakshmanan, L. V. S., & Jagadish, H. V. (2004). Tree Logical Classes for Efficient Evaluation of XQuery. In *ACM SIGMOD International Conference on Management of Data (SIGMOD 04), Paris, France* (pp. 71-82). New York: ACM Press.

Pokorný, J. (2002). XML Data Warehouse: Modelling and Querying. In *5th International Baltic Conference (BalticDB&IS 02), Tallinn, Estonia* (pp. 267-280). Tallinn: Institute of Cybernetics.

Park, B. K., Han, H., & Song, I. Y. (2005). XML-OLAP: A Multidimensional Analysis Framework for XML Warehouses. In *7th International Conference on Data Warehousing and Knowledge Discovery (DaWaK 05), Copenhagen, Denmark* (pp. 32-42). *Lecture Notes in Computer Science,* 3589. Berlin: Springer.

Rajugan, R., Chang, E., & Dillon, T. S. (2005). Conceptual Design of an XML FACT Repository for Dispersed XML Document Warehouses and XML Marts. In *5th International Conference on Computer and Information Technology (CIT 05), Shanghai, China* (pp. 141-149). Los Alamitos: IEEE Computer Society.

Ravat, F., Teste, O., & Zurfluh, G. (2006). Constraint-Based Multi-Dimensional Databases. In Ma, Z. (Ed.), *Database Modeling for Industrial Data Management* (pp. 323-368). Hershey: Idea Group Publishing.

Rusu, L. I., Rahayu, J. W., & Taniar, D. (2005). A Methodology for Building XML Data Warehouses. *International Journal of Data Warehousing & Mining*, 1(2), 67-92.

Thomas, H., & Datta, A. (2001). A Conceptual Model and Algebra for On-Line Analytical Processing in Decision Support Databases. *Information Systems Research*, 12(1), 83-102.

Vrdoljak, B., Banek, M., & Rizzi, S. (2003). Designing Web Warehouses from XML Schemas. In *5th International Conference on Data Warehousing and Knowledge Discovery (DaWaK 03), Prague, Czech Republic* (pp 89-98); *Lecture Notes in Computer Science,* 2737. Berlin: Springer.

Wiwatwattana, N., Jagadish, H. V., Lakshmanan, L. V. S., & Srivastava, D. (2007). X^3: A Cube Operator for XML OLAP. In *23rd International Conference on Data Engineering (ICDE 07), Istanbul, Turkey* (pp. 916-925). Los Alamitos: IEEE Computer Society.

Wang, H., Li, J., He, Z., & Gao, H. (2005). OLAP for XML Data. In *5th International Conference on Computer and Information Technology (CIT 05), Shanghai, China* (pp. 233-237). Los Alamitos: IEEE Computer Society.

Xyleme, L., (2001). Xyleme: A Dynamic Warehouse for XML Data of the Web. In *International Database Engineering & Applications Symposium (IDEAS 01), Grenoble, France* (pp. 3-7). Los Alamitos: IEEE Computer Society.

Zhang, J., Wang, W., Liu, H., & Zhang, S. (2005). X-Warehouse: Building Query Pattern-Driven Data. In *14th international conference on World Wide Web (WWW 05), China, Japan* (pp. 896-897). New York: ACM Press.

## TERMS AND DEFINITIONS

XML data warehouse: a data warehouse managing multidimensionally modeled XML data.

Web warehouse: "a shared information repository. A web warehouse acts as an information server that supports information gathering and provides value added services, such as transcoding, personalization." (Cheng et al., 2000)

XML document warehouse: an XML document repository dedicated to e-business and Web data analysis.

XML-OLAP or XOLAP: OLAP for XML; approaches and operators providing answers to analytical queries that are multidimensional in nature and applied to XML data.

XML graph: data model representing the hierarchical nature of XML data. In an XML graph, nodes represent elements or attributes.

Complex data: data that present several axes of complexity for analysis, e.g., data represented in various formats, diversely structured, from several sources, described through several points of view, and/or versioned.